\documentclass{PoS}
\usepackage{amsmath}

\newcommand{\de}{\ensuremath{\delta} }
\newcommand{\eps}{\ensuremath{\epsilon} }
\newcommand{\la}{\ensuremath{\lambda} }
\newcommand{\si}{\ensuremath{\sigma} }
\newcommand{\cA}{\ensuremath{\mathcal A} }
\newcommand{\cD}{\ensuremath{\mathcal D} }
\newcommand{\cDbar}{\ensuremath{\overline{\mathcal D}} }
\newcommand{\ehat}{\ensuremath{\widehat e} }
\newcommand{\ghat}{\ensuremath{\widehat g} }
\newcommand{\Ibb}{\ensuremath{\mathbb I} }
\newcommand{\cN}{\ensuremath{\mathcal N} }
\newcommand{\cO}{\ensuremath{\mathcal O} }
\newcommand{\cP}{\ensuremath{\mathcal P} }
\newcommand{\cQ}{\ensuremath{\mathcal Q} }
\newcommand{\cU}{\ensuremath{\mathcal U} }
\newcommand{\cUbar}{\ensuremath{\overline{\mathcal U}} }
\newcommand{\lalat}{\ensuremath{\lambda_{\rm lat}} }
\newcommand{\muhat}{\ensuremath{\widehat \mu} }
\newcommand{\UN}{\ensuremath{\mbox{U(}N\mbox{)}} }
\newcommand{\SUN}{\ensuremath{\mbox{SU(}N\mbox{)}} }
\newcommand{\Uone}{\ensuremath{\mbox{U(1)}} }
\renewcommand{\Re}{\ensuremath{\mbox{Re}} }
\renewcommand{\Im}{\ensuremath{\mbox{Im}} }
\newcommand{\X}{\ensuremath{\!\times\!} }
\newcommand{\lra}{\ensuremath{\longrightarrow} }
\newcommand{\Tr}[1]{\ensuremath{\mbox{Tr}\left[ #1 \right]} }
\newcommand{\vev}[1]{\ensuremath{\left\langle #1 \right\rangle} }

\newcommand{\eq}[1]{Eq.~\ref{#1}}
\newcommand{\fig}[1]{Fig.~\ref{#1}}
\newcommand{\refcite}[1]{Ref.~\cite{#1}}

\title{Latest results from lattice $\cN = 4$ super Yang--Mills}
\ShortTitle{Latest results from lattice $\cN = 4$ supersymmetric Yang--Mills}

\author{\speaker{David Schaich} \\
  Institute for Theoretical Physics, University of Bern, 3012 Bern, Switzerland \\
  E-mail: \email{schaich@itp.unibe.ch} \\
}

\author{Simon Catterall \\
  Department of Physics, Syracuse University, Syracuse, New York 13244, United States \\
  E-mail: \email{smcatter@syr.edu} \\
}

\author{Poul H.~Damgaard \\
  Niels Bohr International Academy and Discovery Center, Niels Bohr Institute, \\ University of Copenhagen, 2100 Copenhagen, Denmark \\
  E-mail: \email{phdamg@nbi.dk} \\
}

\author{Joel Giedt \\
  Department of Physics, Applied Physics and Astronomy, \\ Rensselaer Polytechnic Institute, Troy, New York 12065, United States \\
  E-mail: \email{giedtj@rpi.edu}
}

\abstract{ 
  We present some of the latest results from our numerical investigations of $\cN = 4$ supersymmetric Yang--Mills theory formulated on a space-time lattice.
  Based on a construction that exactly preserves a single supersymmetry at non-zero lattice spacing, we recently developed an improved lattice action that is now being employed in large-scale calculations.
  Here we update our studies of the static potential using this new action, also applying tree-level lattice perturbation theory to improve the analysis of the potential itself.
  Considering relatively weak couplings, we obtain results for the Coulomb coefficient that are consistent with continuum perturbation theory.
}

\FullConference{34th annual International Symposium on Lattice Field Theory \\ 24--30 July 2016 \\ University of Southampton, United Kingdom}

\begin{document}
\setlength{\abovedisplayskip}{6 pt}
\setlength{\belowdisplayskip}{6 pt}
Non-perturbative investigations of $\cN = 4$ supersymmetric Yang--Mills (SYM) formulated on a space-time lattice have advanced rapidly in recent years.
In addition to playing important roles in holographic approaches to quantum gravity, investigations of the structure of scattering amplitudes, and the conformal bootstrap program, $\cN = 4$ SYM is also the only known four-dimensional theory for which a lattice regularization can exactly preserve a closed subalgebra of the supersymmetries at non-zero lattice spacing $a > 0$~\cite{Kaplan:2005ta, Unsal:2006qp, Catterall:2007kn, Damgaard:2008pa, Catterall:2009it}.
Based on this lattice construction we have been pursuing large-scale numerical investigations of $\cN = 4$ SYM that can in principle access non-perturbative couplings for arbitrary numbers of colors $N$.
Here we discuss a selection of our latest results from this work in progress.

Last year we introduced a procedure to regulate flat directions in numerical computations by modifying the moduli equations in a way that preserves the single exact supersymmetry at non-zero lattice spacing~\cite{Catterall:2015ira, Schaich:2015daa, Schaich:2015ppr}.
This procedure produces a lattice action that exhibits effective $\cO(a)$ improvement, with significantly reduced discretization artifacts that vanish much more rapidly upon approaching the continuum limit.
We have implemented this improved action in our parallel software for lattice $\cN = 4$ SYM~\cite{Schaich:2014pda}, and are now employing it in the large-scale numerical computations discussed in this proceedings.
We make our software publicly available to encourage independent investigations and the development of a lattice $\cN = 4$ SYM community.\footnote{{\tt\href{http://github.com/daschaich/susy}{http://github.com/daschaich/susy}}}

In this proceedings, after briefly reviewing the improved action we revisit our lattice investigations of the static potential~\cite{Catterall:2014vka, Catterall:2014vga}.
In addition to the new lattice action, we also improve the static potential analysis itself by applying tree-level lattice perturbation theory.
We observe a coulombic potential $V(r) = A - C / r$ and our preliminary results for the Coulomb coefficient $C(\la)$ are consistent with continuum perturbative predictions.
A separate contribution to these proceedings~\cite{Giedt:2016lat} discusses our efforts to investigate S~duality on the Coulomb branch of $\cN = 4$ SYM where some of the adjoint scalar fields acquire non-zero vacuum expectation values leading to spontaneous symmetry breaking.
These efforts involve measuring the masses of the elementary W~boson and the corresponding dual topological 't~Hooft--Polyakov monopole.
\refcite{Giedt:2016lat} also provides an update on our ongoing investigations of the Konishi operator scaling dimension.

\section*{Improved lattice action for $\cN = 4$ SYM} 
Our lattice formulation of $\cN = 4$ SYM is based on the Marcus (or Geometric-Langlands) topological twist of the continuum theory~\cite{Marcus:1995mq, Kapustin:2006pk}.
This produces a $\UN = \SUN \otimes \Uone$ gauge theory with a five-component complexified gauge field $\cA_a$ in four space-time dimensions.
We discretize the theory on the $A_4^*$ lattice, exactly preserving the closed subalgebra $\left\{\cQ, \cQ\right\} = 0$ involving the single twisted-scalar supercharge $\cQ$.
The improved lattice action that we use is~\cite{Catterall:2015ira}
\begin{align*}
  S & = \frac{N}{2\lalat} \sum_n \bigg\{\Tr{\cQ \left(\chi_{ab}(n)\cD_a^{(+)}\cU_b(n) + \eta(n)\left\{\cDbar_a^{(-)}\cU_a(n) + G\cO(n) \Ibb_N\right\} - \frac{1}{2}\eta(n) d(n) \right)} \\
    &                                   -\frac{1}{4} \Tr{\eps_{abcde}\ \chi_{de}(n + \muhat_a + \muhat_b + \muhat_c) \cDbar_c^{(+)} \chi_{ab}(n)} + \mu^2 \sum_a \left(\frac{1}{N} \Tr{\cU_a(n) \cUbar_a(n)} - 1\right)^2\bigg\},
\end{align*}
where the operator \cO in the first line is $\cO = \sum_{a \neq b} \left(\det\cP_{ab} - 1\right)$ and $\cP_{ab} = \cP_{ba}^*$ is the oriented plaquette built from the complexified gauge links $\cU_a$ in the $a$--$b$ plane.
Repeated indices are summed and the forward/backward finite-difference operators $\cD_a^{(\pm)}$ both reduce to the usual covariant derivatives in the continuum limit~\cite{Catterall:2007kn, Damgaard:2008pa}.
All indices run from 1 through 5, corresponding to the five symmetric basis vectors of the four-dimensional $A_4^*$ lattice~\cite{Unsal:2006qp, Catterall:2014vga}. 

When $\mu, G = 0$ this action has the same form as the twisted continuum theory~\cite{Marcus:1995mq, Kapustin:2006pk}.
These two tunable couplings are introduced to stabilize numerical calculations by regulating flat directions and exact zero modes.
The scalar potential with coupling $\mu$ lifts flat directions in the SU($N$) sector, while the plaquette determinant with coupling $G$ does so in the U(1) sector.
Although non-zero $\mu$ softly breaks the \cQ supersymmetry, the plaquette determinant deformation is \cQ exact.
This $\cQ$-exact deformation results from the general procedure introduced in \refcite{Catterall:2015ira}, which imposes the \cQ Ward identity $\sum_n \cO(n) = 0$ by modifying the equations of motion for the auxiliary field,
\begin{equation}
  d(n) = \cDbar_a^{(-)}\cU_a(n) \qquad \lra \qquad d(n) = \cDbar_a^{(-)}\cU_a(n) + G\cO(n) \Ibb_N.
\end{equation}
With $\cO = \sum_{a \neq b} \left(\det\cP_{ab} - 1\right)$ this Ward identity gives $\vev{\Re\det\cP_{ab}} = 1$ after averaging over the lattice volume, while $\vev{\Im\det\cP_{ab}}$ is constrained by the scalar potential.
Thanks to the reduced soft supersymmetry breaking enabled by this procedure, \cQ Ward identity violations vanish $\propto (a / L)^2$ in the continuum limit~\cite{Schaich:2015ppr}.
This is consistent with the $\cO(a)$ improvement expected since \cQ and the other lattice symmetries forbid all dimension-5 operators~\cite{Catterall:2015ira}.

With $\mu, G = 0$ the moduli space of the lattice theory survives to all orders of lattice perturbation theory~\cite{Catterall:2011pd}.
If nonperturbative effects such as instantons also preserve the moduli space, then the most general long-distance effective action $S_{\rm eff}$ contains only the terms in the improved action above~\cite{Catterall:2014vga, Catterall:2014mha}.
In addition, all but one of the coefficients on the terms in $S_{\rm eff}$ can be absorbed by rescaling the fermions and the auxiliary field, leaving only a single coupling that may need to be tuned to recover the full symmetries of $\cN = 4$ SYM in the continuum limit.

\vspace{-6 pt} 
\section*{Tree-level improvement for the lattice $\cN = 4$ SYM static potential} 
\vspace{-6 pt} 
We extract the static potential $V(r)$ from the exponential temporal decay of rectangular Wilson loops $W(\vec n, t) \propto \exp\left[-V(r) t\right]$.
To easily analyze all possible spatial separations $\vec n$ we gauge fix to Coulomb gauge and compute $W(\vec n, t) \equiv \Tr{P(\vec x, t, t_0) P^{\dag}(\vec x + \vec n, t, t_0)}$, where $P(\vec x, t, t_0)$ is the product of complexified temporal links $\cU_t$ at spatial location $\vec x$, extending from timeslice $t_0$ to timeslice $t_0 + t$.

The static potential analysis can be improved by refining the scalar distance $r$ associated with the spatial three-vector $\vec n$.
This is a long-established idea in lattice gauge theory, dating back at least to \refcite{Lang:1982tj}.
Previously we identified the scalar distance as the euclidean norm of $r_{\mu} = \sum_{i = 1}^3 n_i \ehat_{i\mu}$, where each \ehat is a basis vector of the $A_4^*$ lattice.
Because these basis vectors are not orthogonal, $r_{\mu}$ is a four-vector in physical space-time even though $\vec n$ is a three-vector displacement on a fixed timeslice of the lattice.

To obtain tree-level improvement we instead extract the scalar distance $r_I$ from the Fourier transform of the bosonic propagator computed at tree level in lattice perturbation theory.
Then $V(r_I) = 1 / (4\pi r_I)$ to this order in lattice perturbation theory.
Using the tree-level lattice propagator computed in \refcite{Catterall:2011pd}, we have
\begin{equation}
  \label{eq:FT}
  \frac{1}{r_I^2} \equiv 4\pi^2 G(r) = 4\pi^2 \int_{-\pi / a}^{\pi / a} \frac{d^4 k}{(2\pi)^4} \frac{\cos(r\cdot k)}{4\sum_{\mu = 1}^4 \sin^2(k\cdot \ehat_{\mu} / 2)}.
\end{equation}
In this expression $r$ is the same four-vector discussed above while $k_{\mu} = 2\pi \sum_{i = 1}^4 p_i \ghat_{i\mu}$ and the dual basis vectors \ghat are defined by $\ehat_i \cdot \ghat_j = \ehat_{i\mu} \ghat_{j\mu} = \de_{ij}$.
The last identity allows us to replace $r\cdot k = 2\pi \vec n \cdot \vec p$ and $k\cdot \ehat_{\mu} / 2 = \pi p_{\mu}$, more directly relating $r_I$ to the three-vector displacement $\vec n$.

On a finite $L^3\X N_t$ lattice, the continuous integral in \eq{eq:FT} would reduce to a discrete sum over integer $p_{\mu}$.
Since we have not yet computed the zero-mode ($p = 0$) contribution to the discrete sum, here we determine $r_I$ by numerically evaluating the continuous integral that corresponds to the infinite-volume limit.
\refcite{Lang:1982tj} argues that infinite-volume $r_I$ can safely be used in finite-volume lattice calculations, without affecting either the Coulomb coefficient or the string tension.
In agreement with this argument, we checked that both approaches give us similar results even though we currently omit the zero-mode contribution from the finite-volume computation.

We experimented with three integrators to numerically evaluate the four-dimensional integral in \eq{eq:FT}, obtaining consistent results but significantly different performance.
For our problem the most efficient integrator we were able to find was the \verb!Divonne! algorithm implemented in the \verb!Cuba! library~\cite{Hahn:2004fe}.
This is a stratified sampling algorithm based on \verb!CERNLIB! routine D151~\cite{Friedman:1981ed}.
Especially for large $r_I$ \verb!Divonne!'s evaluation of \eq{eq:FT} converged several orders of magnitude more rapidly than two versions of the \verb!vegas! algorithm~\cite{Lepage:1977sw} that we tested.
These two versions of \verb!vegas! both provide some improvements over the original algorithm, and are implemented in \verb!Cuba! and at {\tt\href{http://github.com/gplepage/vegas}{http://github.com/gplepage/vegas}}.

\section*{Latest results for the static potential} 
\begin{figure}[bp]
  \centering
  \includegraphics[width=0.45\linewidth]{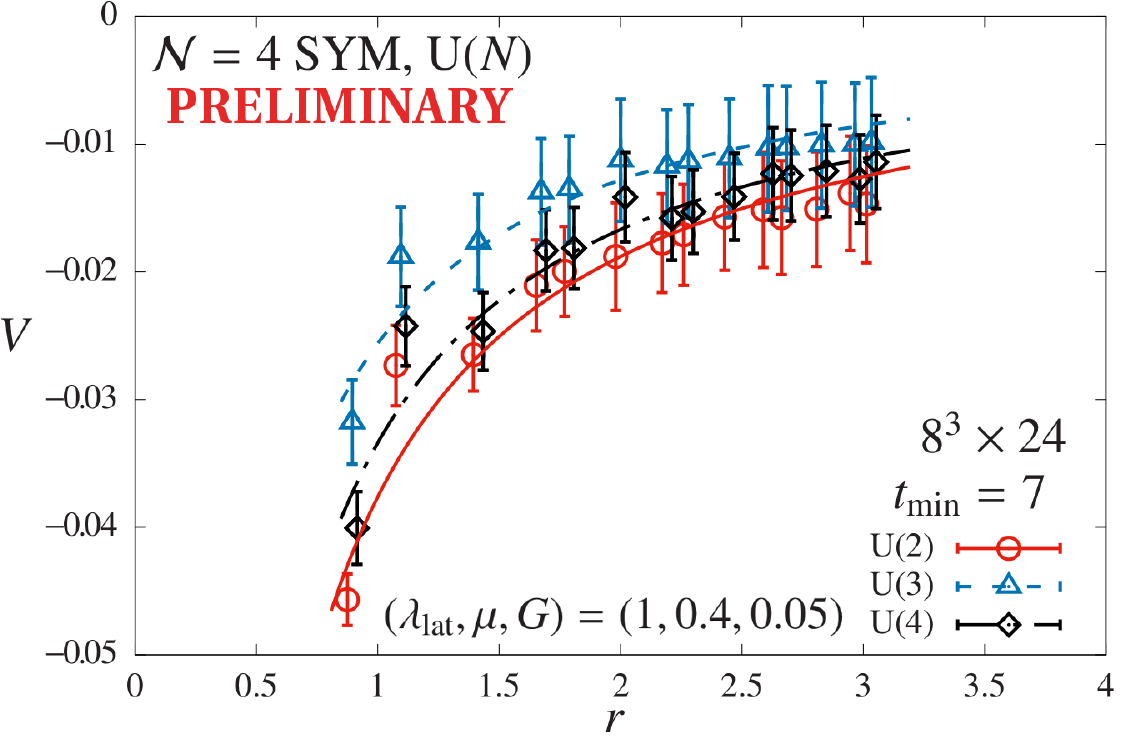}\hfill \includegraphics[width=0.45\linewidth]{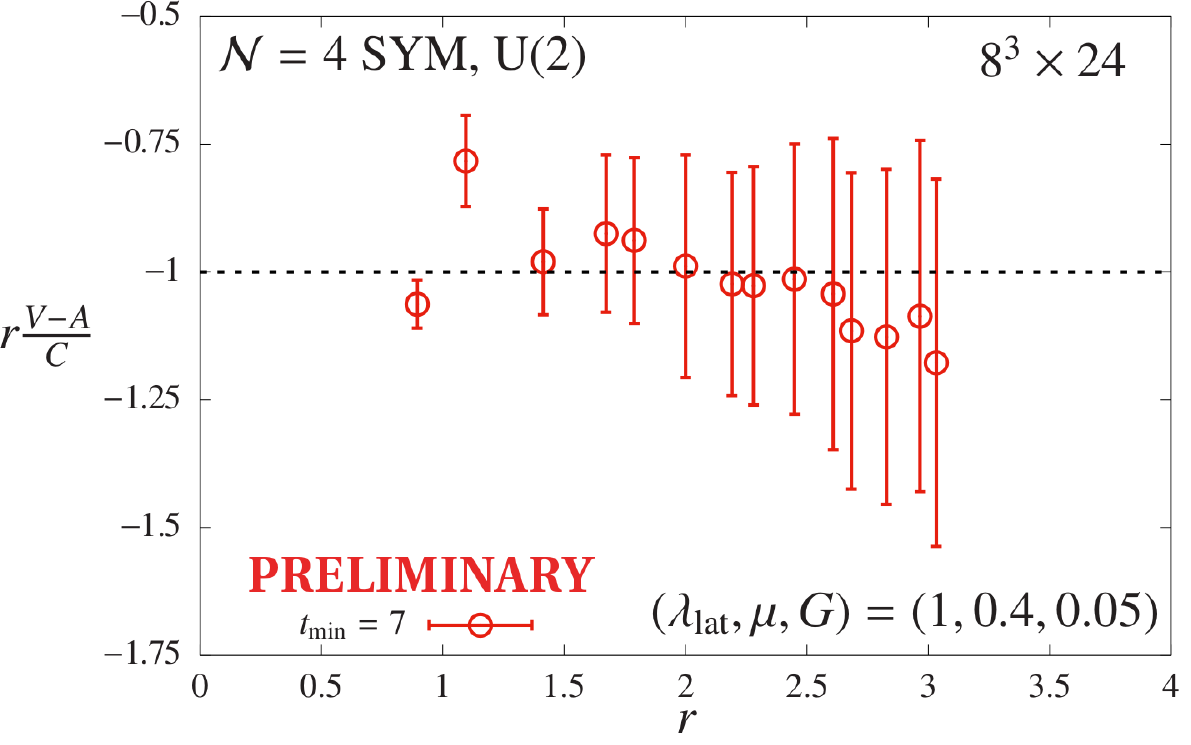} \\[12pt]
  \includegraphics[width=0.45\linewidth]{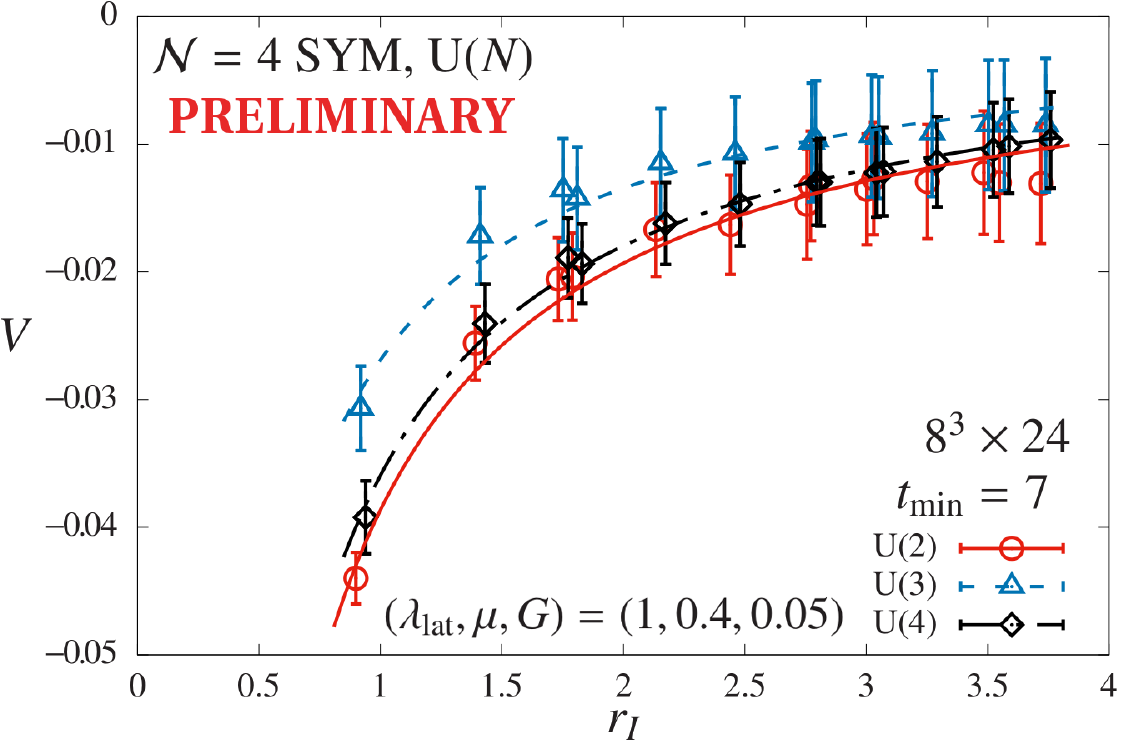}\hfill \includegraphics[width=0.45\linewidth]{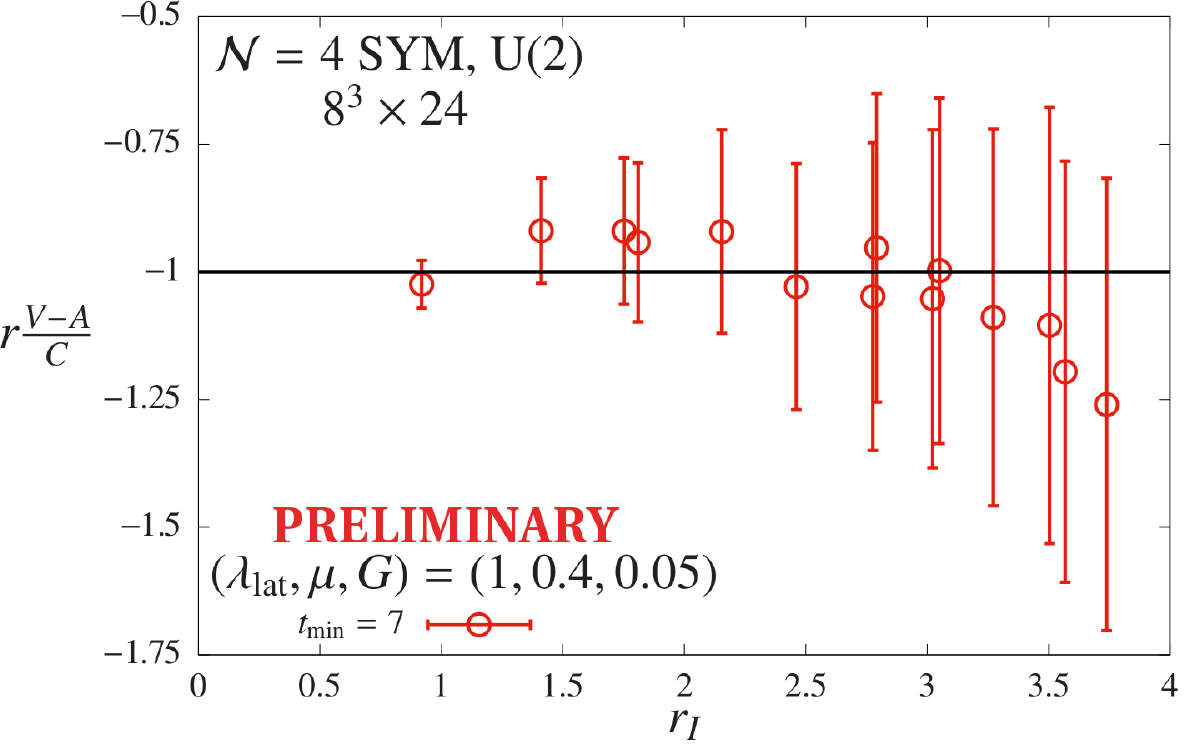}
  \caption{\label{fig:improvement}Unimproved (top) and tree-level improved (bottom) results for the static potential from lattice $\cN = 4$ SYM, on $8^3\X 24$ lattices at 't~Hooft coupling $\la = \lalat / \sqrt{5} = 1 / \sqrt{5}$.  The plots on the left show $V(r)$ for U($N$) gauge groups with $N = 2$ (solid red), 3 (dashed blue) and 4 (dash-dotted black), including fits to the Coulomb form $V(r) = A - C / r$ (all shifted so that $A = 0$).  Those on the right extract the scatter of the $N = 2$ points around the fit by plotting $r\frac{V - A}{C}$.  Tree-level improvement significantly reduces this scatter at short distances, allowing more accurate and reliable analyses.}
\end{figure}

In \fig{fig:improvement} we demonstrate the effects of tree-level improvement for lattice $\cN = 4$ SYM computations of the static potential.
All four plots in this figure consider $8^3\X 24$ lattices generated using the improved action at 't~Hooft coupling $\la = \lalat / \sqrt{5} = 1 / \sqrt{5}$.
The top row of plots analyze $V(r)$ with the scalar distance defined by the naive euclidean norm of $r_{\mu}$.
In the top-left plot we show the potential itself for gauge groups U($N$) with $N = 2$, 3 and 4, including fits to the Coulomb form $V(r) = A - C / r$.
It is possible to see that the first points at $r \approx 0.9$ are consistently below the fit curves, while the next points at $r \approx 1.1$ are well above them.
This scatter of the points around the $N = 2$ fit is isolated in the top-right plot where we show $r\frac{V - A}{C}$ .

It is precisely this scatter at short distances that tree-level improvement ameliorates, as shown in the bottom row of plots.
These results come from the same gauge configurations and measurements as those in the top row, with the only change in the analysis being the use of $r_I$ obtained from \eq{eq:FT} via the \verb!Divonne! integrator in \verb!Cuba!.
There is not a one-to-one correspondence between the points in the two rows of plots.
Several $\vec n$ that produce the same euclidean norm (and are therefore combined in our original analyses) lead to distinct $r_I$.
At the same time, the finite-volume effects also change.
We drop any displacements that extend at least halfway across the spatial volume of the lattice.
When working with euclidean norms for $L = 8$, this imposes $r < r(1, 1, 4) \approx 3.3$, whereas $r_I < r_I(0, 0, 4) \approx 3.8$.

\begin{figure}[bp]
  \centering
  \includegraphics[width=0.45\linewidth]{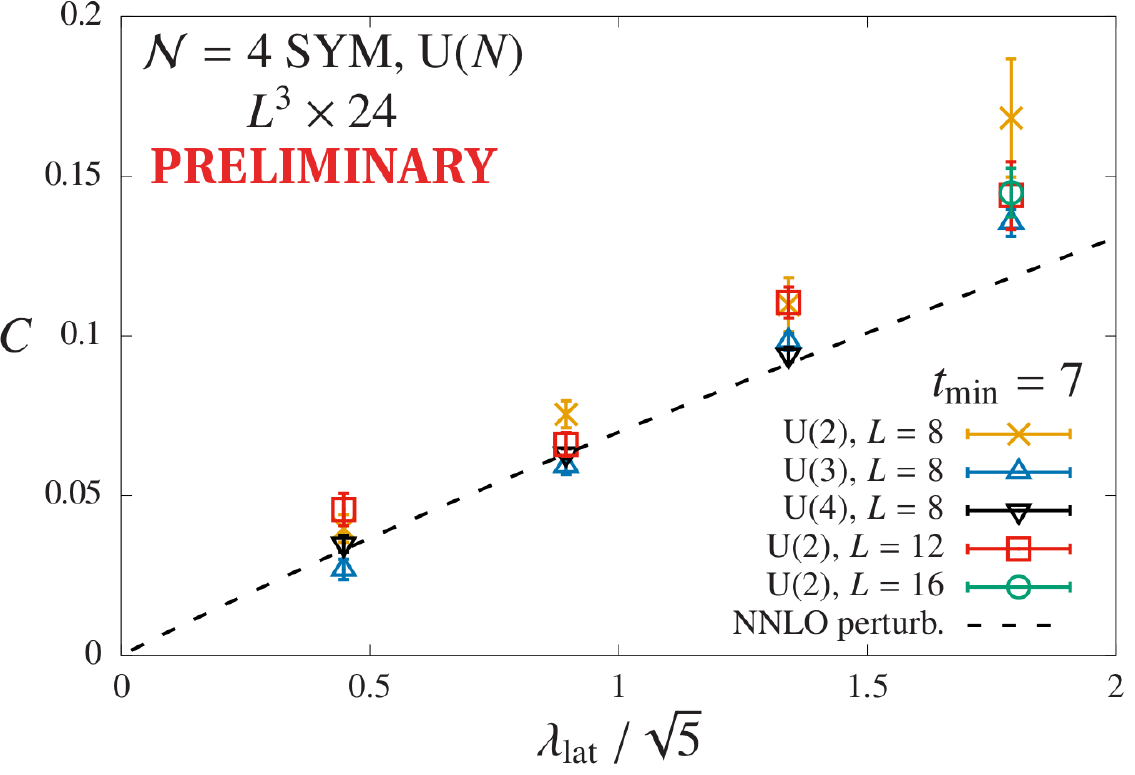}\hfill \includegraphics[width=0.45\linewidth]{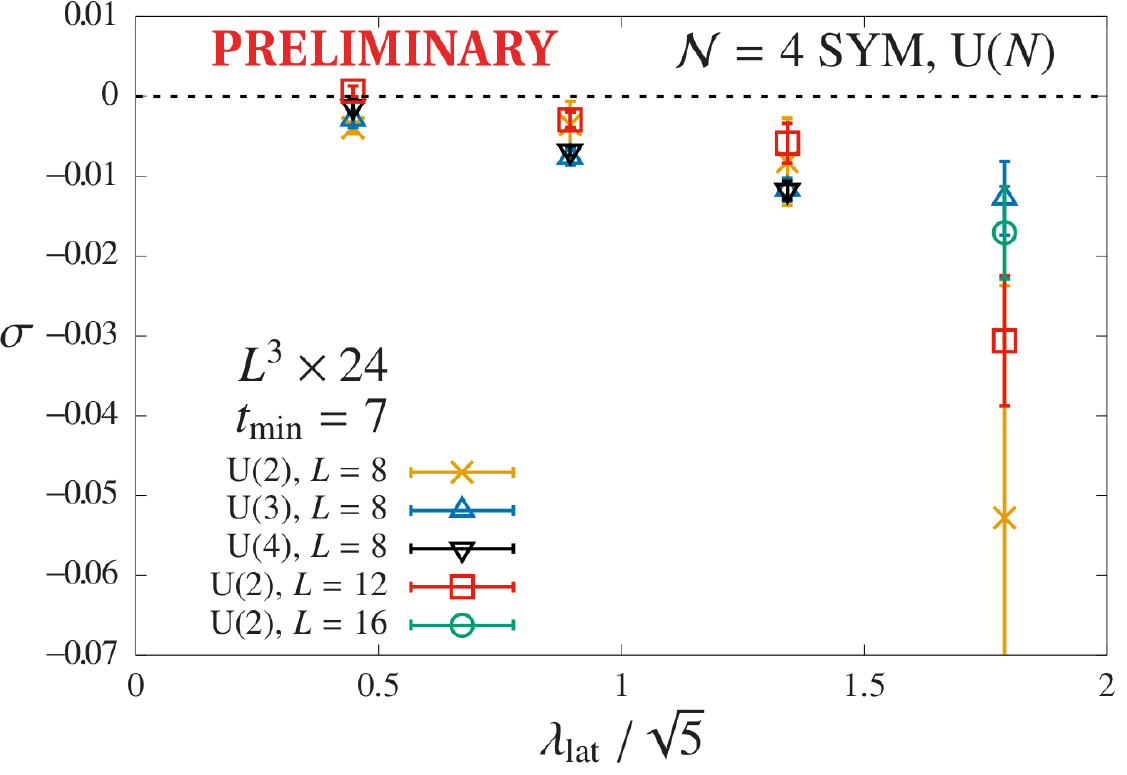}
  \caption{\label{fig:C}Preliminary lattice $\cN = 4$ SYM results from tree-level improved static potential analyses.  {\bf Left:} The Coulomb coefficient $C$ vs.\ the 't~Hooft coupling $\la = \lalat / \sqrt{5}$ on $8^3\X 24$ lattices for U($N$) gauge groups with $N = 2$ (gold $\times$s), 3 (blue triangles) and 4 (black $\nabla$s), along with $N = 2$ results from larger $L^3\X 24$ volumes with $L = 12$ (red squares) and 16 (green circles).  The results are consistent with perturbation theory, and the agreement improves with increasing $N$ and $L$.  {\bf Right:} The string tension \si obtained from the same ensembles upon fitting $V(r) = A - C / r + \si r$.  The small negative results move toward zero as $L$ increases, confirming that the static potential remains coulombic.}
\end{figure}

In \fig{fig:C} we collect preliminary results from tree-level improved static potential analyses employing our new ensembles of gauge configurations generated using the improved action.
On $8^3\X 24$ lattices we consider three U($N$) gauge groups with $N = 2$, 3 and 4, while to explore finite-volume effects we also carry out $12^3\X 24$ and $16^3\X 24$ calculations for $N = 2$.
(Because the larger volumes also help to control discretization artifacts at stronger couplings, so far we have only generated $16^3\X 24$ lattices at the strongest $\la = 4 / \sqrt{5}$ included in this analysis.)
A notable finite-volume effect that we observe is a small negative value for the string tension \si when we fit the static potential to the confining form $V(r) = A - C / r + \si r$.
We can see in \fig{fig:improvement} that such a negative string tension would improve the fit for distances near the finite-volume cutoff.
As $L$ increases we gain data at larger distances, which more effectively constrain $\si$.
In the right plot of \fig{fig:C} we see that the string tension moves toward zero as $L$ increases, confirming that the static potential is coulombic at all couplings we consider.

We therefore fit the static potential to the Coulomb form $V(r) = A - C / r$ to obtain the results for the Coulomb coefficient $C$ in the left plot of \fig{fig:C}.
For the same gauge groups and lattice volumes discussed above our results are consistent with the next-to-next-to-leading-order (NNLO) perturbative prediction from Refs.~\cite{Pineda:2007kz, Stahlhofen:2012zx, Prausa:2013qva}.
The agreement with perturbation theory tends to improve as $N$ and $L$ increase, especially at the strongest 't~Hooft coupling $\la = 4 / \sqrt{5}$ where the larger volume helps control discretization artifacts.

\section*{Next steps for lattice $\cN = 4$ SYM} 
We are near to finalizing and publishing our tree-level improved analyses of the lattice $\cN = 4$ SYM static potential based on the improved lattice action introduced last year and summarized above.
In addition we are making progress analyzing the anomalous dimension of the Konishi operator, developing a variational method to disentangle the Konishi and supergravity ($20'$) operators as described in \refcite{Giedt:2016lat}.
We continue to investigate the possible sign problem of the lattice theory, as well as the restoration of the other supersymmetries $\cQ_a$ and $\cQ_{ab}$ in the continuum limit.
Finally, \refcite{Giedt:2016lat} also presents a new project to study S~duality on the Coulomb branch of the theory, by measuring the masses of the W~boson and the corresponding dual topological 't~Hooft--Polyakov monopole.
Ideally this Coulomb branch investigation will allow non-perturbative lattice tests of S~duality even at 't~Hooft couplings relatively far from the self-dual point $\lalat = 2\pi N\sqrt{5} \approx 14N$.

\vspace{12 pt}
\noindent {\sc Acknowledgments:}~We thank Tom DeGrand, Julius Kuti and Rainer Sommer for helpful discussions of perturbative improvement for the static potential, and Rudi Rahn for advice on numerical integration. 
This work was supported by the U.S.~Department of Energy (DOE), Office of Science, Office of High Energy Physics, under Award Numbers DE-SC0009998 (DS, SC) and DE-SC0013496 (JG). 
Numerical calculations were carried out on the HEP-TH cluster at the University of Colorado, the DOE-funded USQCD facilities at Fermilab, and the Comet cluster at the San Diego Computing Center through the Extreme Science and Engineering Discovery Environment (XSEDE) supported by U.S.~National Science Foundation grant number ACI-1053575. 

\bibliographystyle{utphys}
\bibliography{lattice16}
\end{document}